\documentclass{aa501}
\usepackage{graphics,amssymb,natbib,rotating,psfrag,epsfig,array}

\title{Kerr black holes and time profiles of gamma ray bursts}
\titlerunning{Kerr black holes and time profiles of gamma ray bursts}

\author{S.\,McBreen\inst{1} \and
        B.\,McBreen\inst{1} \and
        L.\,Hanlon\inst{1} \and
        F.\,Quilligan\inst{1,2}}

\institute{Department of Experimental Physics, University College
Dublin, Dublin 4, Ireland \and Intel Corporation, Leixlip, Co.
Kildare, Ireland}

\offprints{smbreen@bermuda.ucd.ie}
\date{Received / Accepted}

\abstract{The cumulative light curves of Gamma Ray Bursts (GRBs)
smooth the spiky nature of the running light curve.  The
cumulative count increases in an approximately linear way with time
t for most bursts.  In 19 out of 398 GRBs with T$_{90}
>$ 2 s, the cumulative light curve was found to increase with
time as $\sim$ t$^{2}$ implying a linear increase in the running
light curve . The non-linear sections last for a substantial
fraction of the GRB duration, have a large proportion of the
cumulative count and many resolved pulses that usually end with
the highest pulse in the burst.  The reverse behaviour was found
in 11 GRBs where the running light curve decreased with time and
some bursts are good mirror images of the increases. These GRBs
are among the spectrally hardest bursts observed by BATSE. The
most likely interpretation is that these effects are signatures
of black holes that are either being spun up or down in the
accretion process. In the spin up case, the increasing Kerr
parameter of the black hole allows additional rotational and
accretion energy to become available for extraction. The process
is reversed if the black hole is spun down by magnetic field
torques. The luminosity changes in GRBs are consistent with the
predictions of the BZ process and neutrino annihilation and thus
provide the link to spinning black holes. GRBs provide a new
window for studying the general relativistic effects of Kerr
black holes.
}

\begin{document}
 \maketitle

\section{Introduction}

It is suspected that spinning black holes reside in a variety of
astrophysical sources.  Frame dragging creates a special region
called the ergosphere in which any material or energy must rotate
in the same direction as the black hole.  The energetic reactions
near the black hole maybe responsible for relativistic jets in
active galactic nuclei (AGN), microquasars and gamma ray bursts
\citep{koide:2002,frail:2001}. In GRBs the source of the enormous
energy in gamma rays maybe the cataclysmic formation of a
spinning black hole involving mergers of compact objects such as
neutron star (NS) binaries or NS and black holes
\citep{piran:1999,ruffjan:1999} and also during or after the
collapse of massive stars
\citep{macfad:1999,pacy:1998,vietri:1998,reeves:2002}. The central
engine is hidden from view and only gravitational radiation and
neutrinos may escape and reach the observer directly from the
engine.  A key feature of the internal shock model is that the
observed gamma rays reflect the variability of the central engine
and the GRB duration may be determined by the engine
\citep{reemes:1994,piran:1999}. The cumulative output in gamma
rays of a burst indirectly reflects the output of the central
engine via a relativistic jet.  The advantage of
using the cumulative light curve is that it reveals the trends by
smoothing the spiky nature of the running light curve.  The
cumulative light curves of most bursts can be approximated by a
linear function of time and GRBs may be regarded as relaxation
systems that continuously accumulate energy in the reservoir and
discontinuously release it \citep{mcbreenb:2002}. In a relatively
small number of GRBs, the cumulative light curves depart from
linearity in a consistent way.  The selection and properties of
these GRBs are presented here in sections 2 and 3, discussed in
section 4 and summarised in section 5.

\section{Analysis of the light curves of Gamma Ray Bursts}
A large sample of the brightest BATSE GRBs was used with the data
combined from the four energy channels \citep{fishman:1995}.  The
analysis procedures are described in detail elsewhere (Quilligan
et al., 2002; McBreen et al., 2001 \& 2002a). The full sample
consisted of 100 GRBs with duration T$_{90} <$ 2 s where T$_{90}$
measures the burst integrated count from 5\% to 95\% of the
total, 319 GRBs with T$_{90} >$ 2 s and a further fainter sample
of 79 GRBs with T$_{90} >$ 100 s to include very long bursts. The
cumulative light curves of most GRBs could be approximated by a
linear function implying constant output over most of the
duration of the burst \citep{mcbreenb:2002}. Two significant
minorities were visually identified that are better described by
nonlinear changes in the cumulative count. In category A the
running count increased towards the tallest pulse in the burst
resulting in a nonlinear increase in the cumulative profile. In
category B the running count decreased after the tallest pulse in
the burst causing the cumulative profile to increase at a much
slower rate as time progressed. In category A the normalised
cumulative profile was fit by the function
\begin{equation}
\textrm{R}(t_{\textrm{i}})=\textrm{I}_{\textrm{min}} + c (
t_{\textrm{i}} - t_{\textrm{0}})^\textrm{$\beta$}
\end{equation}
where $\textrm{R}(t_{\textrm{i}})$ is the cumulative count at time
t$_{\textrm{i}}$ and I$_{\textrm{min}}$ is the minimum count at
t$_{\textrm{0}}$. In category B the function used was
\begin{equation}
\textrm{D}(t_{\textrm{i}})=\textrm{I}_{\textrm{max}} - c (
t_{\textrm{max}} - t_{\textrm{i}})^\textrm{$\beta$}
\end{equation}
where $\textrm{D}(t_{\textrm{i}})$ is the cumulative count at time
t$_{\textrm{i}}$, t$_{\textrm{max}}$ is the time at the end of
the fitted section of the burst and I$_{\textrm{max}}$ is the
value of the count at t$_{\textrm{max}}$. I$_{\textrm{max}}$
usually has a value close to unity. The profiles were fit using
the Levenberg-Marquardt nonlinear minimisation algorithm. The
fitted sections include at least 5 pulses $\geq 5 \sigma$ from
the running profile, 20$\%$ of the GRB duration and 30$\%$ of the
cumulative total. In the two categories there were 19 and 11 GRBs
that satisfied the selection criteria. An increase in 
number occurs when the selection criteria are relaxed.

The median number of pulses N in GRBs with T$_{90} > $2 s is only
6 \citep{quilligan:2002} and the requirement on N restricts the
GRBs to about half of the total. The number of pulses is
important because they may originate from explosions in the central
engine and discriminate against GRBs, where the emission is not
well resolved into pulses because of the washing out of time
structure in the jet before the gamma-ray photosphere, and
interactions with the external medium including the effects of
early afterglow \citep{derm:1999}. The changes in the cumulative
count presented here are quite different from the smooth power law
decays in GRBs of the FRED (i.e. Fast Rise Exponential Decay) type
\citep{giblin:2002} and the smooth decays in a sample of pulses
within GRBs \citep{ryde:2001}.

Most GRBs with T$_{90} <$ 2 s can also be approximated by a
linear fit to the cumulative light curve \citep{mcbreenb:2002}.
The GRBs have a median value of N = 2.5 \citep{sheila:2001} and no short GRBs were
found to meet the criteria used for long GRBs.

\section{Results on the light curves of GRBs}

The 19/11 GRBs that satisfied the selection for categories A/B
are listed in Table 1.  The running and cumulative light curves
and fits to the nonlinear sections of the bursts are given in
Fig. 1 and 2 for three bursts in each category.  The cumulative
light curve smoothes the spiky nature of the running light curve.
In category A the highest peak flux frequently occurs near the
end of the nonlinear increase and subsequently the bursts stops
abruptly (Fig. 1a) or continues at a much reduced rate (Fig.
1c) due to decrease in accretion.  In category B there is a decrease in the running light
curve that is an approximate mirror image of bursts in category A
(Fig. 2) and the nonlinear slowdown in the cumulative light curve
results from the drop in amplitude of the pulses in the
running profiles.

The GRBs in Table 1 have large numbers of pulses.  The errors due
to the BATSE counting statistics can be neglected because they
are very small compared with the large fluctuations caused by the
pulses in the running light curves.  Equations 1 and 2 were fit
to the cumulative light curve to obtain the best fit to the data
(Figs. 1 and 2).  The values of $\beta$ are in the range 1.8 to
2.3 (Table 1).  The range in $\beta$ was estimated by varying
$\beta$ from 1.5 to 2.5 and visually estimating the range where
there was agreement between the fitted functions and the
cumulative light curves. The values are listed in Table 1 and
typically are in the range $\pm$0.2. The fluctuations between the
fitted curve and the cumulative data are due to the pulses and
the time intervals between them in the running light curve.  These
residuals are inherent to the pulsed nature of GRBs.  The value
of c (Table 1) is given for the parabolic value $\beta$ = 2 for all bursts to enable
comparison of the normalised cumulative profiles.  The
errors on the hardness ratios (Table 1) are all less than 7\% 
with the exception of the soft GRB 6903 where it is
50\%.

\begin{figure}[ht]
\leavevmode \psfrag{Counts}[t]{\Large Count }
 \psfrag{Cumulative Counts}[t]{\Large Cumulative Count }
  \psfrag{Normalised Fit}[t]{}
\psfrag{Time}[t]{\Large Time (secs) }
\psfrag{Time (secs)}[t]{\Large Time (secs) }
\psfrag{Relative Time}[t]{\large Relative Time (secs)}
\begin{center}
\resizebox{0.9\columnwidth}{0.18\textheight}{\includegraphics{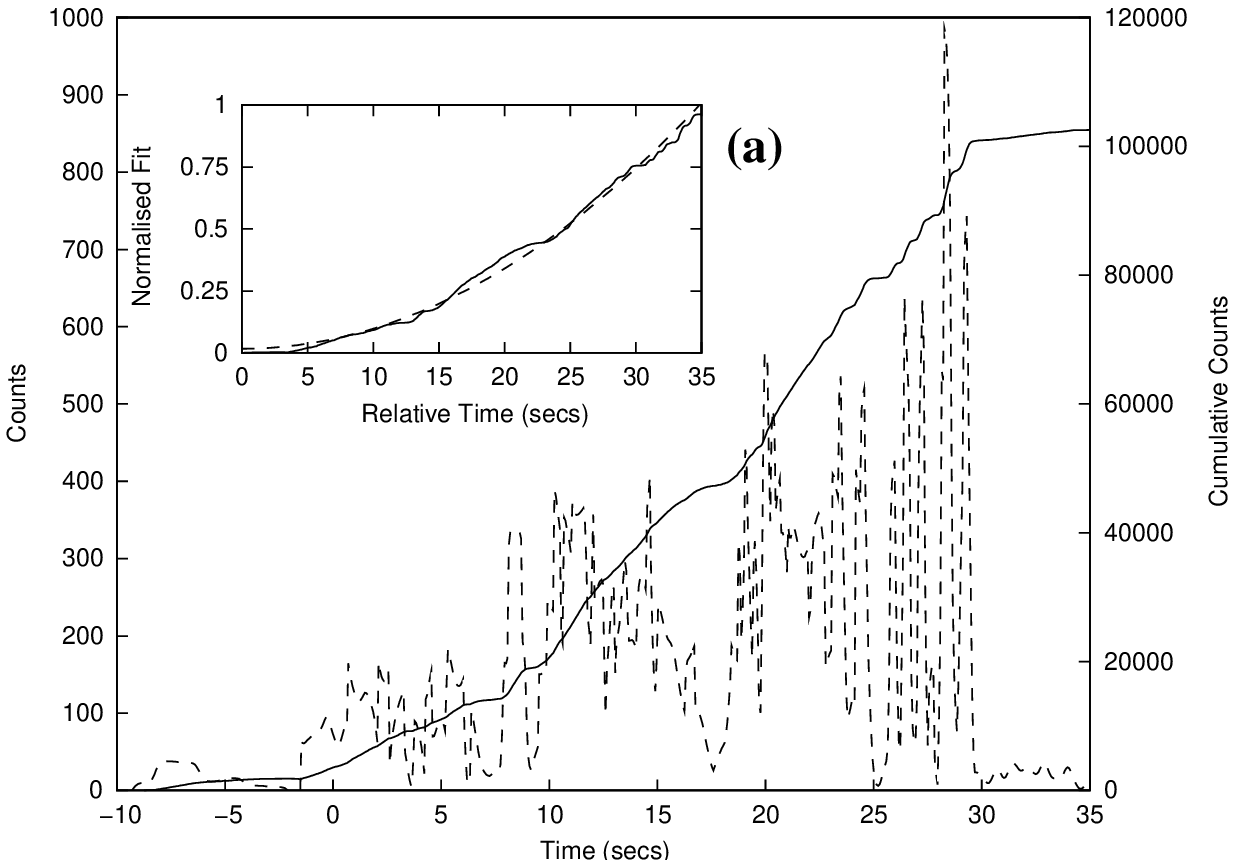}}\\[9pt]
\resizebox{0.9\columnwidth}{0.18\textheight}{\includegraphics{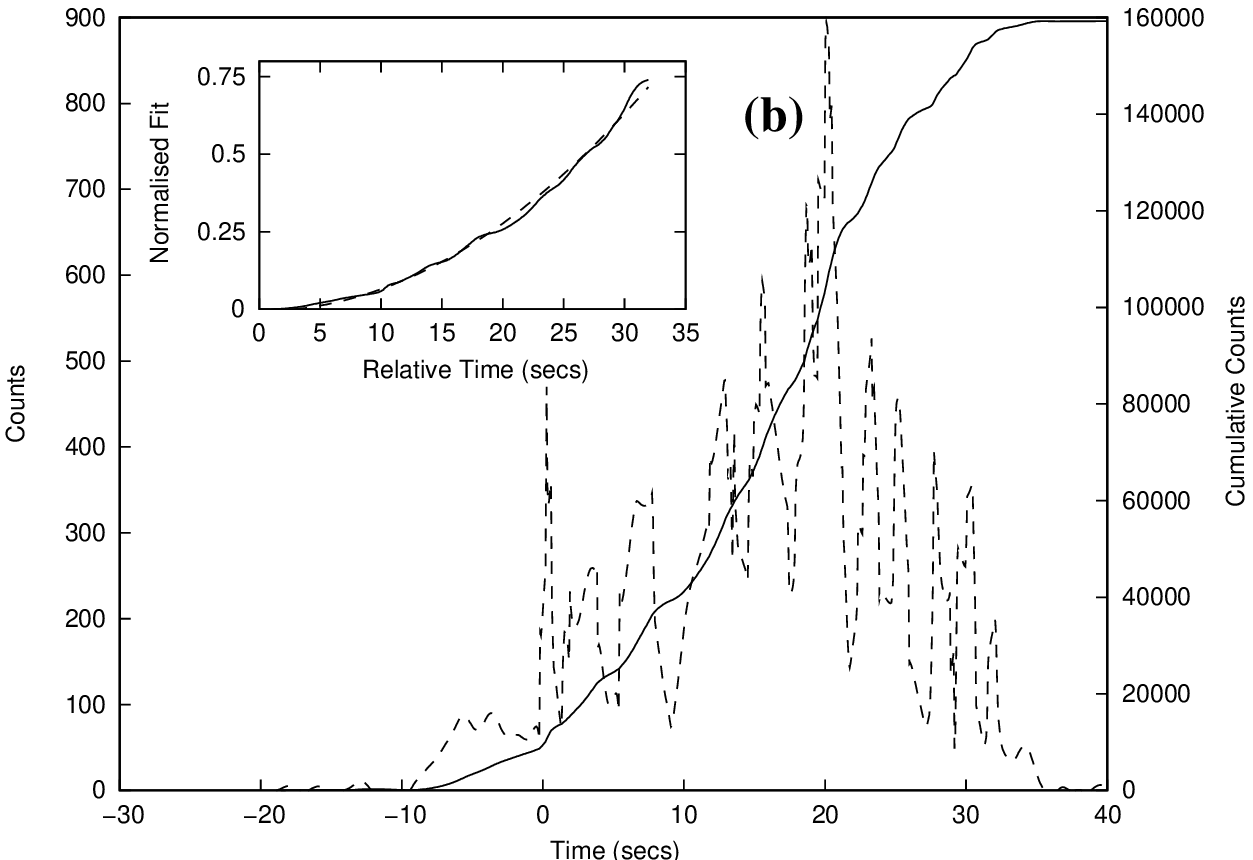}}\\[9pt]
\resizebox{0.9\columnwidth}{0.18\textheight}{\includegraphics{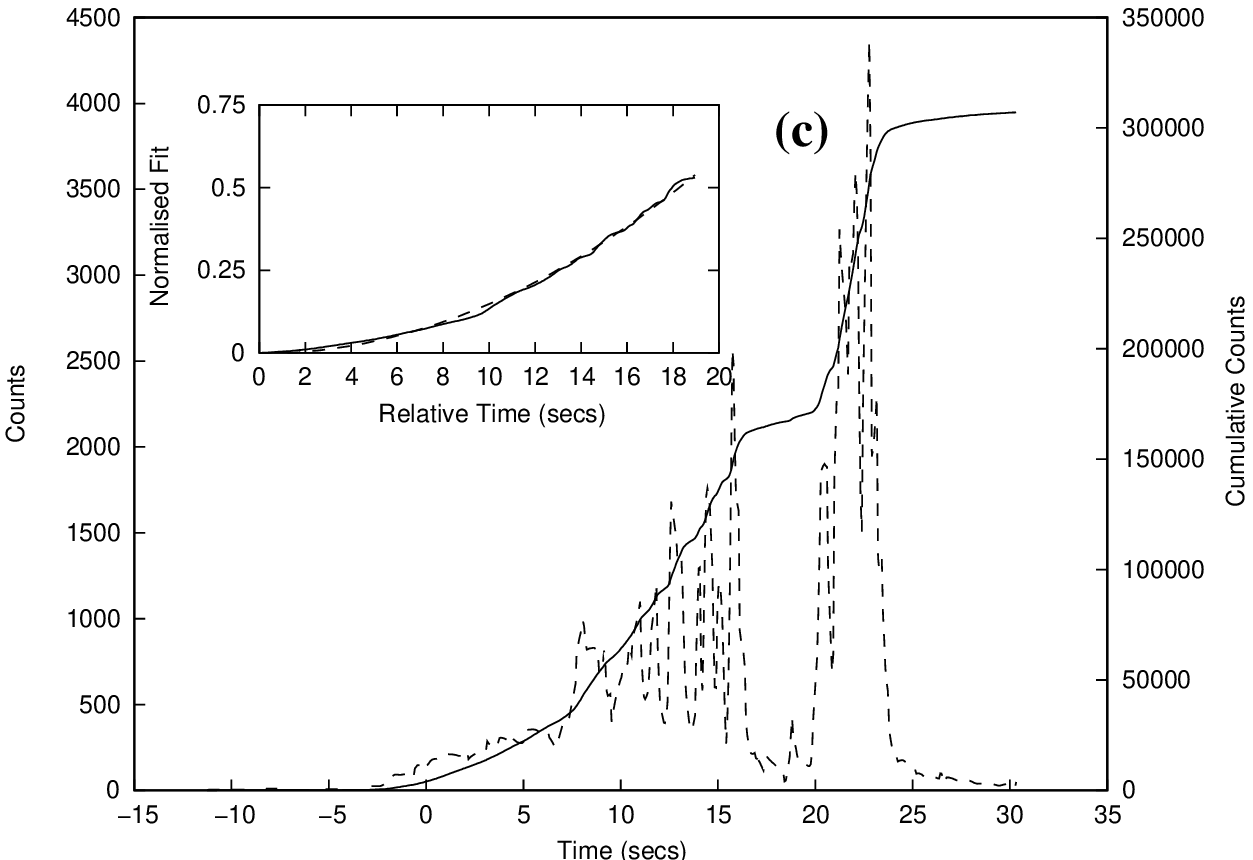}}
 \caption{The running (dashed) and cumulative (solid) light
    curves of the BATSE bursts with trigger numbers a) 3105, b)
    3860 and c) 6963 with count per 64 ms and cumulative count scales on
    the left and right vertical axes.  The inserts give the
    cumulative count (solid) and the fit of the function (dashed) for the
    relevent section. The start and end times are listed in Table
    1 and t$_{1}$ is shifted to zero for the inserts.
    The vertical axes in the inserts are the normalised cumulative
    count.}
 \end{center}
\end{figure}

\begin{figure}[t]
 \leavevmode \psfrag{Counts}[t]{\Large
Count }
 \psfrag{Cumulative Counts}[t]{\Large Cumulative Count }
 \psfrag{Normalised Fit}[t]{}
\psfrag{Time}[t]{\Large Time (secs) }
\psfrag{Relative Time}[c]{\large Relative Time (secs)}
\begin{center}
\resizebox{0.9\columnwidth}{0.18\textheight}{\includegraphics{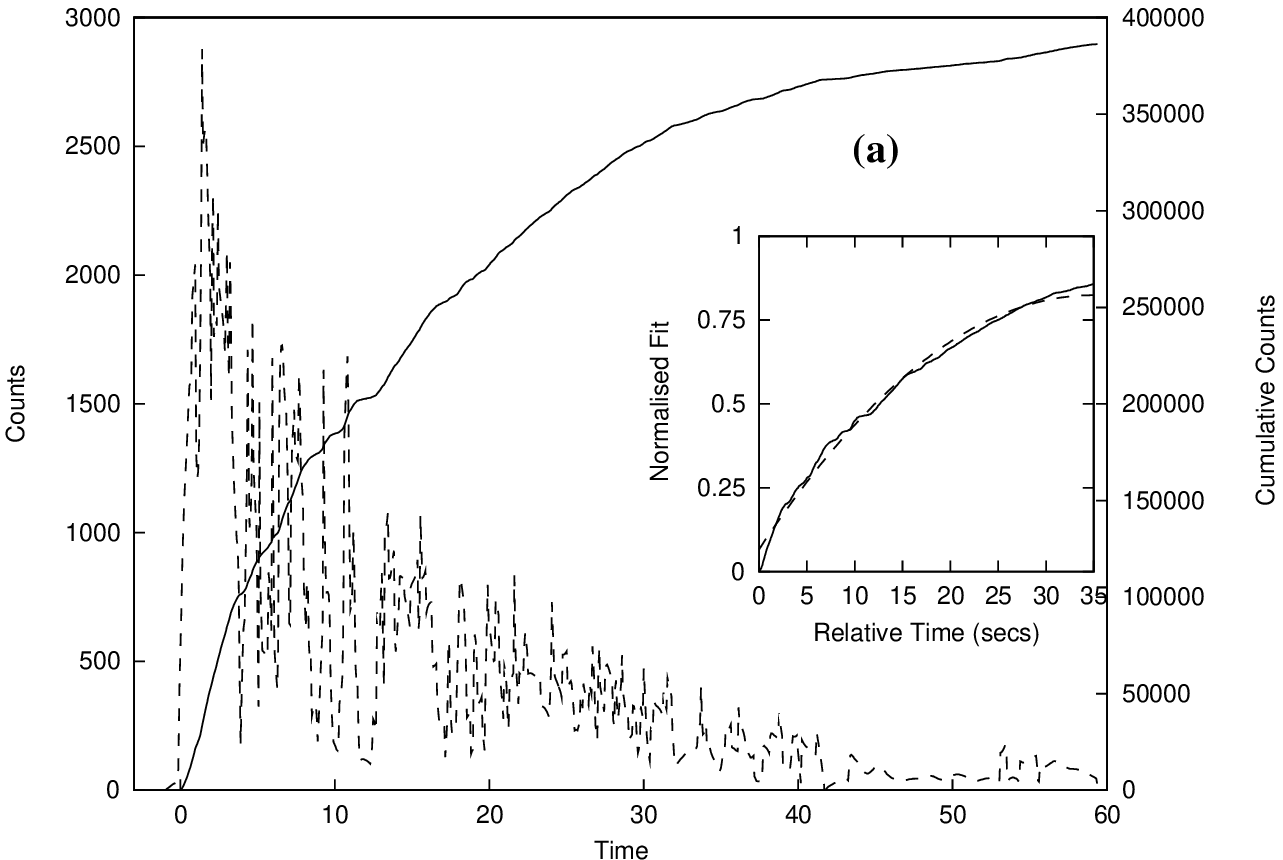}}\\[9pt]
\resizebox{0.9\columnwidth}{0.18\textheight}{\includegraphics{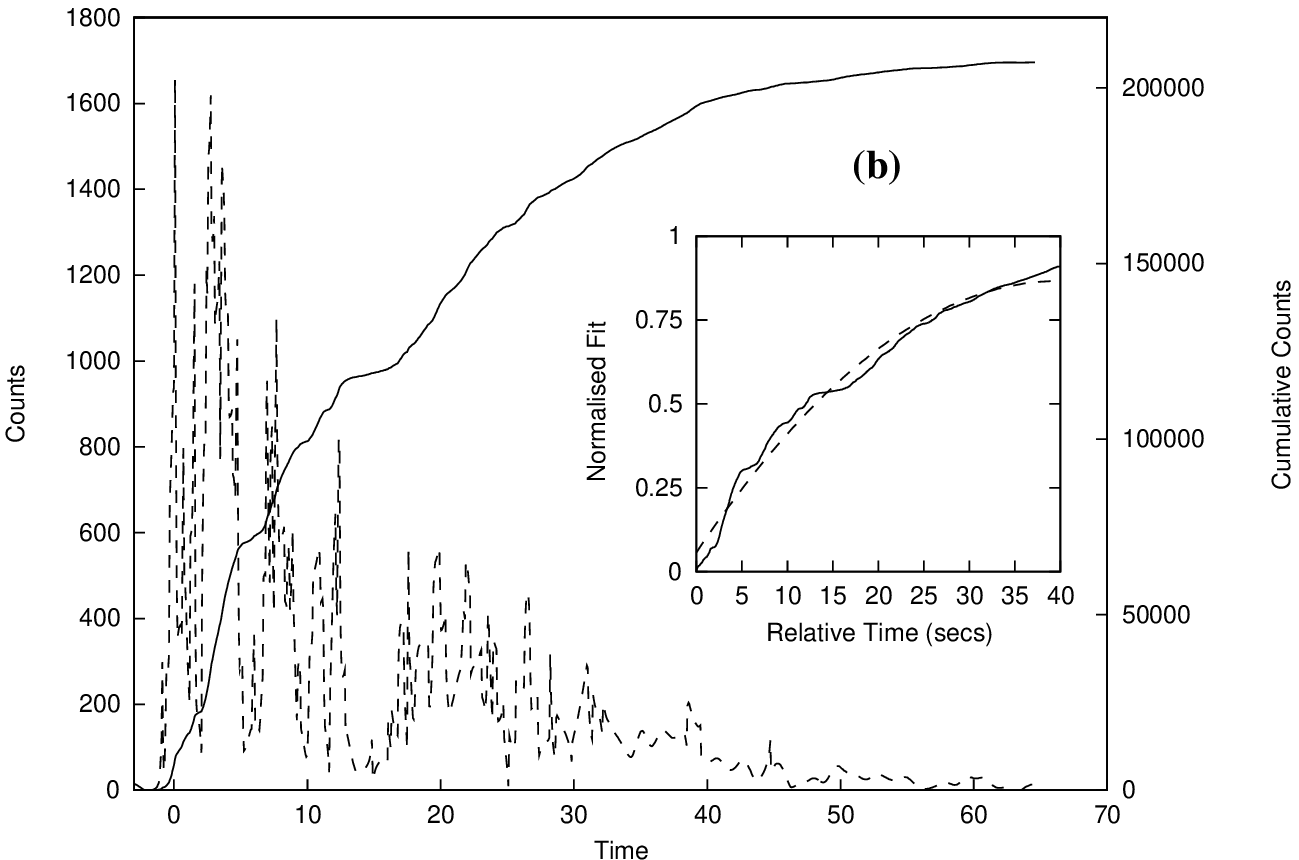}}\\[9pt]
\resizebox{0.9\columnwidth}{0.18\textheight}{\includegraphics{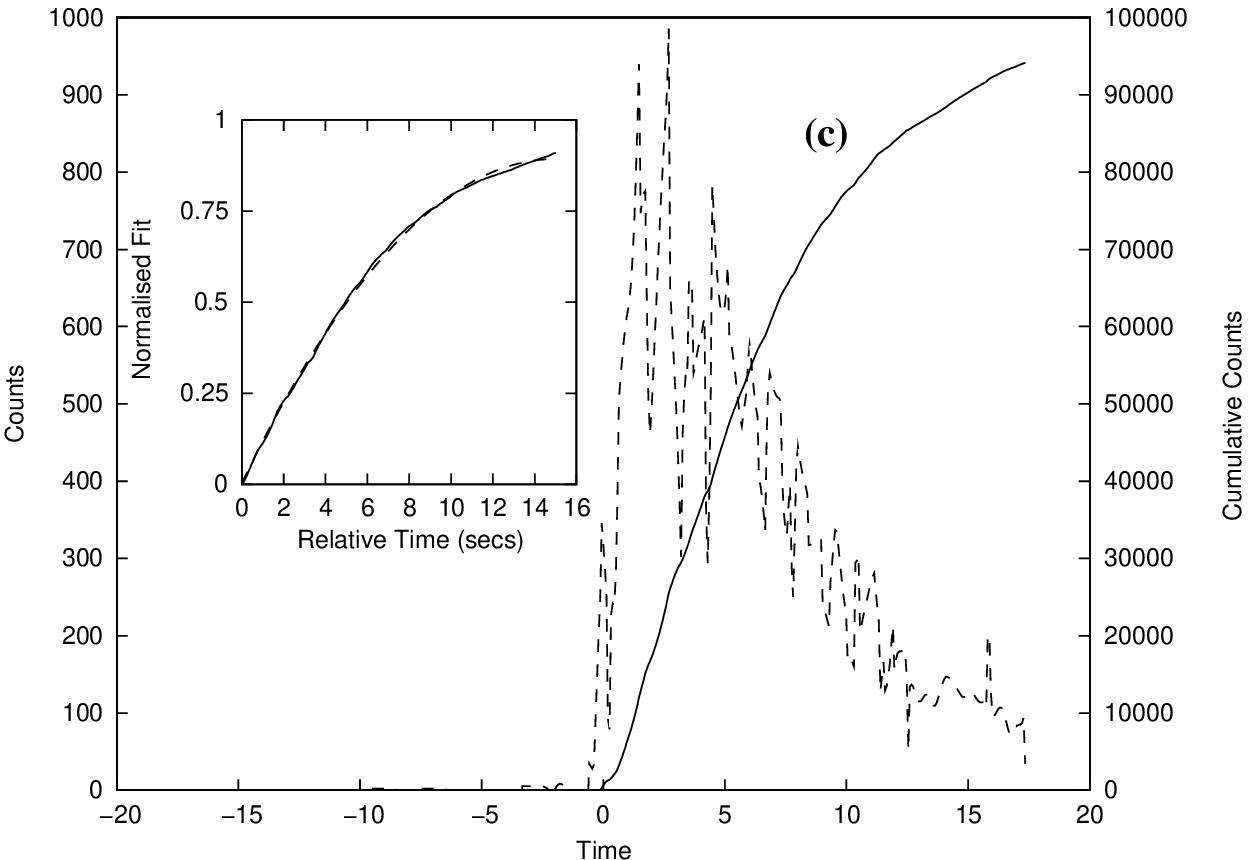}}

    \caption{The running (dashed) and cumulative (solid) light
    curves of the BATSE bursts with trigger numbers a) 678, b)
    4039 and c) 6694 with the same notation as Fig. 1.}
    \end{center}
\end{figure}

\begin{table}[h]
 \caption{
 The GRBs that satisfied the criteria for category A (first 19 rows)
 and category B (second 11 rows).  The columns refer to BATSE
trigger number, the number of pulses N $\geq$ 5 $\sigma$, the
number of those pulses N$^{\prime}$ included in the fitted region
between times t$_{1}$ and t${_2}$ with a minimum value of 7 s for
GRB 1440 and maximum of 140 s for GRB 8101, the hardness ratio
(HR) which is the ratio of fluence in BATSE channels 3 and 4
above 100 keV to that in channels 1 and 2 below 100 keV, the
percentage of the total integrated counts (\%C) in the fitted
region, the index \textrm{$\beta$} and the coefficient $c \times 10^{-4}$ for the
fit with $\beta$ = 2.
GRBs 3247, 6353, 6903, 8101 and 7660 are from the
fainter sample with T$_{90} >$ 100 s.}

\begin{tabular}[b]{lccccccc}
\hline \hline \vspace{0.1cm} Burst   & N/N$^{\prime}$ &
t$_{1}$,t$_{2}$&HR&  $\%$ C & \textrm{$\beta$} & c
\\ \hline
394 & 25/12 & 0,25 &4.6& 42 & 1.9$\pm$0.2 & 6.8  \\
1122  & 10/8 & 0,11&3.2 & 62& 1.9$\pm$0.2 & 49.0 \\
1440&  15/9& 10,17 & 5.3& 70 & 2.1$\pm$0.2& 147.1 \\
2450& 22/11& 45,65& 3.4 & 49& 2.1$\pm$0.3& 12.0   \\
3035 & 21/12& 5,75 &4.7   & 52 & 2.1$\pm$0.3 & 1.1\\
3105 & 30/30& $-$10,30& 6.5 & 98 & 2.1$\pm$0.2 & 6.3\\
3247& 25/17& 80,180& 7.7 & 79 & 2.1$\pm$0.2 & 0.9 \\
3489& 12/5&    $-$2,11 & 6.6& 56 &  2.0$\pm$0.2 & 35.4\\
3860& 13/8 & $-$5,22 & 22.4  & 72 & 1.9$\pm$0.3 & 9.6\\
5526&  34/14& $-$6,16 & 4.4&  36  &  2.1$\pm$0.2& 7.6 \\
6353& 9/5& $-$10,50   & 1.7& 53 & 2.2$\pm$0.2 & 1.5 \\
6453 & 25/12 & $-$5,52 & 1.7 & 46 & 2.0$\pm$0.3 & 1.4\\
6587& 28/20& 3,25  &  7.6& 73 & 2.1$\pm$0.2& 15.8\\
6593 & 20/11& $-$10,14 & 4.7 & 60 & 2.1$\pm$0.3 & 10.2\\
6903& 13/5& $-$25,15 & 1.6&  47& 2.1$\pm$0.2 & 3.3\\
6963 & 17/10 & 0,17 & 4.1& 52 & 2.0$\pm$0.2 & 18.4 \\
7318  & 12/7 &  $-$3,10 & 21.8& 78 & 2.0$\pm$0.2 & 48.0\\
7575& 30/20 & 135,165 & 12.9& 78 & 2.1$\pm$0.2 & 9.1\\
8101& 11/9& $-$40,100  & 10.4& 95 & 1.9$\pm$0.2 & 0.5\\ \hline
678 & 52/44& 1,36   & 42.6& 86 & 1.9$\pm$0.2 & 6.2\\
2891& 18/14& 0.5,20   & 23.1& 89 & 2.1$\pm$0.3 & 22.0 \\
2929 & 44/36& 12,50 & 13.5& 83 & 2.2$\pm$0.3 & 5.7\\
2984 & 22/12& 6,20  & 14.2& 72 & 2.1$\pm$0.2 & 37.2\\
2993 & 17/11 & 0,20 & 31.1 &75 & 2.2$\pm$0.2 & 16.4\\
2994 & 36/22& 6,35  & 19.3& 58 & 2.3$\pm$0.2 & 6.3\\
3408 & 44/35& 10,60 & 6.4& 77 & 1.9$\pm$0.2 & 3.2\\
4039 & 33/31& 0,40 & 21.5& 91 & 1.9$\pm$0.3 & 5.1\\
6694 & 10/9& 1,16  & 18.3& 91 & 2.1$\pm$0.2 & 39.5\\
7660 & 7/5& 25,140  & 8.4 & 77 & 2.0$\pm$0.3 & 0.7\\
7766 & 22/11 & 0,20& 48.2& 85& 2.0$\pm$0.2 & 18.1 \\ \hline \hline
\end{tabular}
\end{table}

\section{Discussion}

The nonlinear changes in the light curves are usually evident in
all four BATSE channels. The achromatic nature of the signals
seem to exclude an effect due to changes in the optical depth.  It
has been suggested that some of the time structure in GRB light
curves could be due to precession of the black hole equator and
the accretion disk
\citep{portegies:1999,fargion:1999,romero:1999,thorne3:1986,macfad:1999}.
The estimated precessional period is about 1 second and is too
small to account for the gradual increases and decreases in the
GRB light curves. The nonlinearities in the light curves could be
a subtle signature of the central engine that only occur in some
GRBs. Most models of GRBs favour a hyper-accretion process into a
newly formed black hole and in these systems the nonlinearities
might be due to systematic changes in the accretion rate or to
general relativistic effects of a spinning black hole.  All the
evolutionary scenarios, such as mergers of compact objects and
collapsars, create rapidly rotating black holes because they
accrete a fraction of their mass from a disk.  However the engine
must be capable of generating GRBs with many pulses with
progressively increasing amplitudes and GRBs that appear to be
their mirror images with harder spectra.

When a black hole swallows matter and radiation from an accretion
disk, its mass and angular momentum evolve.  The evolution of the
black hole was first evaluated by \citet{bardeen:1970} following
the initial work of \citet{lynden:1969}.  The spin of the black
hole, as measured by the Kerr parameter {\it a} = Jc/GM$^{2}$
where M is the mass of the black hole and J its angular momentum,
increases nonlinearily with accretion from 0 to approach but not
equal 1 for an extreme Kerr black hole \citep{thorne:1974}.  The
equatorial radius of the ergosphere decreases from $r_{e}$ =
2GM/c$^{2}$ to GM/c$^{2}$ and the radius of the marginally stable
orbit also decreases from $r_{ms}$ = 6GM/c$^{2}$ to GM/c$^{2}$
and additional gravitational potential energy becomes available.
The dependance of {\it a} on accreted mass is given by
\citet{thorne:1974} and plotted in \citet{popham:1999}.
Substantial accretion is required to significantly alter the
value of {\it a} which acquires a value $a \sim$ 0.97 when the
mass of the hole has been accreted.  Many models of GRBs require
extremely high accretion rates and for a typical long duration
GRB could be about 0.1 M\(_{\odot}s^{-1}\).  There could be
substantial changes in {\it a} during a GRB and it is necessary
to know how the power output varies with {\it a} and time.
Assuming a constant mass accretion rate {\it a} evolves with time
as $\sim$ t$^{0.5}$ for 0.3 $< a <$ 0.8 and more slowly/rapidly
for $a >$ 0.8/$ a <$ 0.3 \citep{popham:1999}.

There are three energy extraction processes to consider and precise
predictions are not always available because of the complexity of
these systems \citep{mesz:2000,rees:1999}.  In the standard accretion
process, the viscous energy dissipation is balanced by neutrino
cooling.  \citet{popham:1999} called it neutrino dominated
accretion flow (NDAF) and worked out its properties including
variations in the disk mass, accretion rate, black hole mass and
Kerr parameter.  The properties of these disks have been further
explored \citep{narayan:2001}.  One extraction mechanism is the
\(\nu \bar{\nu} \rightarrow e^{+}e^{-}\) process that taps the
thermal energy of the disk.  The annihilation of neutrinos close
to the axis of a rotating system, where there is a low
concentration of baryons, was proposed as a way to produce a
clean fireball \citep{elp:1989,meszaros:1992}.  A number of models
have been used to estimate the neutrino emission from NS binaries
and NS and black holes
\citep{ruffjan:1999,leew:2000,janka:1999,rosswog:2002}. 
In a more complete model \citep{popham:1999} the
neutrino luminosity and energy deposition in the polar regions
was numerically evaluated using the Kerr metric.  The integrated
energy deposited by \(\nu \bar{\nu} \rightarrow e^{+}e^{-}\) increases with time as
$\sim t^{2}$ when {\it a} increases above an initial value of 0.5
(Figs. 21 and 22 in \citet{macfad:1999}) and including neutrinos
down to the last stable orbit.  The calculations reveal that at
higher accretion rates the neutrino luminosity and efficiency for
annihilation increase.  These parameters also increase with the
Kerr parameter {\it a}.  As {\it a} increases the last stable
orbit moves inward.  The neutrino emission from the higher
gravitational potential increases both the luminosity and
temperature and also increases the density of neutrinos because
of the more compact geometry \citep{asano:2001,salmonson:1999}.
The model predictions are in good agreement with the cumulative
light curves presented here.  However the annihilation rate
scales as the square of the neutrino luminosity \citep{mesz:2000}
and the efficiency is low if the production is too gradual.

An alternative and more efficient process for extracting energy
maybe dissipation of magnetic fields generated by differential
rotation in the disk.  This extraction process is largely
independent of {\it a} and includes a
relativistic wind or jet from the disk and flares 
\citep[e.g.][]{kluzrud:1998,rees:1999,livio:1999}.
However a large
energy source is also available in the spin of the black hole that
maybe extracted by MHD coupling in the Blandford-Znajek (BZ)
process \citep{blandford:1977} and the MHD Penrose process
\citep{koide:2002}.  The rotational energy available is 0.29
Mc$^{2}$ for a maximally rotating black hole.  The BZ process
behaves as an interaction of the black hole with its surrounding
electromagnetic field and depends directly on the spin of the
black hole.
The predicted luminosity is L \(\sim 10^{50}\,
a^{2}\)(M/3$M_{\odot}$)$^{2}$(B/10$^{15}$G)$^{2}$ ergs s$^{-1}$ where the magnetic
field is supplied by the torus. It is interesting that L
\(\propto~a^{2} \propto\) t provided B remains constant on the
horizon. This process also predicts a cumulative luminosity that
increases $\sim$ t$^{2}$.  Furthermore there are the GRBs with
many pulses with progressively declining amplitudes (Fig. 2 and
Table 1) and running light curves that are reasonable mirror
images of the increases in output.  These GRBs appear to be good
examples of the reverse process, namely, the decrease in the Kerr
parameter {\it a} produced by the slowing down of rapidly spinning
black holes by the BZ or other processes
\citep{vanput:2001,leekim:2002,Li:2002}. The BZ process can account for
the behaviour of the cumulative light curves of GRBs by spin-up
and spin-down of the black hole.  

There are significant spectral differences between the bursts in
Table 1 with the median values of the HRs of 5.3 and 19.3 for the
two categories. The remainder of the GRBs with T$_{90} >$ 2 s
have median values of the HR of 4.1.  The corresponding median
values of the peak energies E$_{\rm peak}$ for available GRBs are
340 keV and 956 keV for categories A and B \citep{lloyd:2002}. The
GRBs in category B are among the spectrally hardest BATSE GRBs,
with the value of E$_{\rm peak}$ generally following the intensity
during the burst \citep{pebs:2000}.  The property of the central
engine that maybe responsible for this behaviour is the spin of
the black hole. The BZ mechanism operates most efficiently for a
spun up black hole that is threaded by a powerful magnetic
field.  It appears that this system will generate
more energetic outbursts with spectrally harder GRBs and 
observed range of pulse properties 
\citep[e.g.][]{norris:2002,mcbreenb:2002,lbp:2000,ramfen:2000,quilligan:2002,nakar:2002}. The
progenitor systems should have high angular momentum and also
form smaller mass black holes that are spun up and down more
easily.  The high spin of the black hole and angular momentum
increasing outwards can stabilise the disk against runaway radial
instabilities \citep{daimoc:1998}.

The majority of GRBs with T$_{90} > $ 2 s have cumulative light
curves that increase in a linear way with t. They do not have
substantial nonlinear behaviour that meet the stringent criteria
adopted here. There are a range of possibilities to account for
this steady state behaviour including 
variations in the
accretion rate or magnetic field that mask the slow changes due
to \textit{a}, different progenitors and disk emission dominating
neutrino annihilation and the BZ process. In the later case the majority
of GRBs are powered by emission from the accretion disk.
However the relative
contribution from MHD processes in the disk and spinning black
hole are a matter of debate \citep{leewijers:2000}.

In a few cases there is a gap in the light curves of the
nonlinear increases and decreases where the output appears to be
suspended for a short interval (e.g. Fig. 1c) and BATSE trigger
numbers 2929, 3408 and 8101.  The time intervals between pulses
can be accommodated in models of GRBs as relaxation systems
\citep{rm:2001,mcbreenb:2002}.   GRBs with T$_{90} < $ 2 s are
well described by linear increases in the cumulative counts and
no convincing cases were found with nonlinear behaviour perhaps
because of the short duration of the emission and small number of
pulses.
\section{Conclusions}
The results of this study of BATSE light curve of GRBs are
presented.  The analysis revealed that the running light curve of
a minority of bursts increase or decrease approximately linearly
with time.  These results are interpreted as possible evidence
for spin-up and spin-down of black holes during the burst.  The BZ
process and neutrino annihilation are consistent with these
results.  The brute force appearance of Kerr black holes being
spun up an down by hyperaccretion in GRBs contrasts with the more
sedate and longer lasting indications of black holes in
MCG-6-30-15 and other sources
\citep{wilms:2001,miller:2002,woo:2002,koide:2002}.
\small
\bibliography{katmonic,allrefs_sheila}
\bibliographystyle{apj}
\end{document}